\documentclass[12pt]{article}
\begin{document}

{\LARGE \bf{Software and Sociology in UK Astronomy 

Kevin A.\ Pimbblet}} \\
E-mail:K.A.Pimbblet@durham.ac.uk \\
Address: Department of Physics, University of Durham, South Road, Durham, DH1 3LE, UK.\\
Accepted for publication in Astronomy \& Geophysics.

\section*{Abstract}
I discuss the remit of Starlink's software strategy groups and a
particular item on the agenda of a meeting of the image processing 
software strategy group held on 26$^{\rm{th}}$ January 2001: `Why don't 
people use Starlink software?'.  The answer to this question was 
speculated to be primarily a sociological effect: those people 
supervising UK astronomy Ph.D. students  are largely people who 
had learnt their trade at a time when Starlink had a less than 
perfect reputation.  I report on the recommendations made to Starlink to 
counter this effect.

\section*{Introduction}
In the United Kingdom, the Starlink Project (see http://www.starlink.rl.ac.uk)
has provided interactive data processing facilities for astronomers for
over 18 years.  Starlink has provided both hardware and software on behalf
of PPARC; `the primary purpose \dots [being] \dots to maximize the return 
on PPARC's expenditure on astronomical computing'
(http://www.starlink.rl.ac.uk).  In 1999, Starlink's budget stood at 
approximately $2.5$ million pounds sterling per annum.  
Out of this sum, approximately $0.6$ million pounds is ear-marked for
the provision of software.  These provisions include the development, 
distribution and documentation of the software.
  
The Starlink software collection is a set of software written both by 
members of the Starlink Project and imported from outside the Project 
for the use of the UK astronomical community.  Many utilities and packages 
exist, mostly focussing on data reduction but there are also some (limited) 
theoretical tools.

\section*{Software Strategy Groups}
A mainstay of Starlink's software development is the existence of Software 
Strategy Groups (SSGs).  There are six such SSGs whose remit covers 
different aspects of astronomical research:
spectroscopy, image processing\footnote{The Author of this article has 
been a member of the Image Processing SSG since 1999.}, information 
services \& databases, radio, mm, \& sub-mm astronomy, theory \& 
infrastructure and X-ray astronomy.  
The SSGs were set up after the 1991/1992 review of Starlink to provide
feedback, re-assess software priorities and identify future objectives.
Membership of the SSGs is composed of both Starlink's programmers and 
expert users drawn from across a broad spectrum of personal experience and 
opinion.
The SSGs are annually asked to advise on strategies to be adopted over 
in the coming years in order to `satisfy (and continue to satisfy) the 
needs identified by users in the [Starlink Users] questionnaire. \dots
They will also be asked to identify in more detail those 
software projects that should be carried out over the next year in order to
implement the strategy'
(http://star-www.rl.ac.uk/star/docs/sgp44.htx/node3.html).

\section*{Sociological Issues}
Third on the agenda of the image processing SSG for $26^{\rm{th}}$ January 
2001 was the loaded question `Why don't people use Starlink software?'
This question came as a surprise for a few members of the SSG as many of 
the users used Starlink's software on a daily basis in the research.  

Starlink software, however, is by no means the only available source of 
software for astronomers.  The major competition for users is the  
successful Image Reduction and Analysis Facility (IRAF).  IRAF is a publicly 
available `general purpose software system for the reduction and analysis 
of scientific data' 
(http://star-www.rl.ac.uk/iraf/web/faq/FAQsec01.html)\footnote{IRAF written 
and supported by the IRAF programming group at the National Optical
Astronomy Observatories (NOAO) in Tucson, Arizona.}.  IRAF overlaps 
many of the areas that the Starlink software does (e.g. general purpose 
image reduction software).  One may expect that there would be 
substantial competition for users between the two, however, this is not 
the case.  UK astronomers have \emph{tended} in the past to migrate
towards using the IRAF option in preference to Starlink's software.
Contributory factors to this may be that in many observatories (e.g. the 
Isaac Newton Group) IRAF has historically been much more readily 
available than Starlink and UK astronomers have used IRAF to be 
more compatible with their American collaborators. 
So why do many UK astronomers use IRAF when much of the equivalent Starlink 
software is arguably better, more stable and more reliable?  
In answering this question, the image processing SSG expediently pointed 
out that this was not always the case.  During the 1980's, Starlink 
software had a less than perfect reputation.  
But why in 2001 is there still such a trend for astronomers to be 
preferentially using IRAF packages over Starlink ones?
The image processing SSG speculated that the major reason was a 
sociological one.
Post-graduate students who are presently commencing work on their Ph.D. 
theses will primarily be introduced to the tools of their trade by their 
Ph.D. supervisors.  Their supervisors, on the whole, in turn learnt their 
trade during, or lived through the aforementioned time when Starlink 
had a less than perfect reputation.  
Their supervisors prejudices and perspectives naturally influence their 
students: 
in guiding their post-graduate students they are likely to recommend 
IRAF in preference to Starlink software to their charges.  
Thus, Starlink software has become less used over time, potentially 
creating a further problem: a lack of local expert users.  
Even if new post-graduate students chose to use Starlink software, they 
would probably have little to no personal guidance available to them 
whereas if they were to chose IRAF, many people within the given astronomy 
department would potentially be able to assist.

\section*{Discussion}
To remedy this situation, the image processing SSG has recommended a 
number of changes be implemented.  Firstly, to give the software a higher 
visibility, the Starlink newsletter would be re-introduced to publicize 
new software developments.  More ideas were to implement a method for 
disseminating what software is available on-line; the introduction of a 
`man' page; and an education scheme to provide novice and infrequent users 
reach rapid decisions about what Starlink software could be helpful.
Although these recommendations are a start, it is probably going to take 
some time to wash away the memories of the imperfect reputation that 
Starlink had in the 1980s and counter the present sociological prejudice 
against Starlink software.  
In the immediate future, many new software products will be released.  
Of particular note is the Interactive Data Language (IDL).  
The recent rise and increasing use of IDL has shown that quality, new 
software will always have a place within the astronomical community and 
Starlink must re-double its efforts to compete.

\section*{Acknowledgements}
I sincerely acknowledge the data analysis facilities provided by the 
Starlink Project which is run by CCLRC on behalf of PPARC, without which 
my Ph.D. would not be what it is.  I thank Peter Draper and Nigel 
Metcalfe for useful discussions on the factual content of this article
and the image processing SSG that inspired this.

\end{document}